\documentclass[twocolumn,oldversion]{aa}
\usepackage{graphicx}
\usepackage{txfonts}
\newcommand{\ApJ}{Astrophys. J.}

\newcommand{\PRL}{Phys. Rev. Lett.}
\newcommand{\PRD}{Phys. Rev. D}
\newcommand{\MNRAS}{Mon. Not. Roy. Astron. Soc.}

\newcommand{\AsPhys}{Astropart. Phys.}
\newcommand{\NewAst}{New Astron. }
\newcommand{\Nat}{Nature }
\newcommand{\MNRASL}{Mon. Not. Roy. Astron. Soc. Lett.}
\newcommand{\JCAP}{J. Cosm. Astropart. Phys.}
\newcommand{\AJS}{Astrophys. J. Supp. }
\newcommand{\PASP}{Publ. Astron. Soc. Pac.}
\newcommand{\SAL}{Soviet. Astron. Lett. (Tr:Pisma)}

\long\def\comment#1{}

\def\la{\hbox{ \raise.35ex\rlap{$<$}\lower.6ex\hbox{$\sim$}\ }}
\def\ga{\hbox{ \raise.35ex\rlap{$>$}\lower.6ex\hbox{$\sim$}\ }}

\def\W2{{\cal W}}

\newcommand{\START}{{\rm i}}
\newcommand{\END}{{\infty}}

\newcommand{\rad}{r}

\newcommand{\isw}{{\rm ISW}}
\newcommand{\X}{{\rm G}}

\newcommand{\sky}{{\rm sky}}

\begin{document}

\title{
Optimising large galaxy surveys for ISW detection}

\titlerunning{}

   \author{Marian Douspis\inst{1}, Patricia G. Castro\inst{2}, Chiara Caprini\inst{3}, Nabila Aghanim\inst{1}
         }


  \institute{IAS CNRS, B\^at. 121, Universit\'e Paris-Sud, F-91405 Orsay, France\\
\email{marian.douspis@ias.u-psud.fr, nabila.aghanim@ias.u-psud.fr}
\and
CENTRA, Departamento de F\'{\i}sica,
  Edif\'{\i}cio Ci\^{e}ncia, Piso 4,
  Instituto Superior T\'ecnico,
  Av. Rovisco Pais 1, 1049-001 Lisboa, Portugal.
\email{pgcastro@ist.utl.pt}
\and
IPhT, CEA-Saclay, 91191 Gif-sur-Yvette, France \\
\email{chiara.caprini@cea.fr}
}

\date{}

\abstract{We report on investigations of the power of next generation
  cosmic microwave background and large scale structure surveys in
  constraining the nature of dark energy through the cross-correlation
  of the Integrated Sachs Wolfe effect and the galaxy distribution.
  First we employ a signal to noise analysis to find the most
  appropriate properties of a survey in order to detect the correlated
  signal at a level of more than $4$ sigma: such a survey should cover
  more than $35$\% of the sky, the galaxy distribution should be
  probed with a median redshift higher than $0.8$, and the number of
  galaxies detected should be higher than a few per squared arcmin. We
  consider the forthcoming surveys DUNE, LSST, SNAP, PanSTARRS.  We
  then compute the constraints that the DUNE survey can put on the
  nature of dark energy (through different parametrizations of its
  equation of state) with a standard Fisher matrix analysis. We
  confirm that, with respect to pure CMB constraints,
  cross-correlation constraints help in breaking degeneracies among
  the dark energy and the cosmological parameters.  Naturally, the
  constraining capability is not independent of the choice of the dark
  energy model. Despite being weaker than some other probes (like
  Gravitational Weak-Lensing), these constraints are complementary to
  them, being sensitive to the high-redshift behaviour of the dark
  energy.  \keywords{Cosmology -- Cosmic microwave background -- Large
    scale structure -- Cosmological parameters} } \maketitle
%

\section{Introduction}
\label{sec:introduction}

Measurements of the Cosmic Microwave Background (CMB) angular power
spectra are now invaluable observables for constraining cosmology. The
detailed shape of these spectra allows one to determine cosmological
parameters with high precision. The ``concordance'' model, built over
the years with CMB, Supernovae of type Ia, and galaxy distribution
observations, seems to reproduce quite well most of the cosmological
observables. This model needs the existence of a dark energy component
that may be accounted for by a cosmological constant $\Lambda$.
However, more complex scenarios cannot be ruled out by present
datasets. Among them, one could think of a dark energy component with
an equation of state $w$ different from $w=-1$ (cosmological constant)
or even varying in time $w(z)$ such as in scalar field Quintessence
models.  Moreover, the effect of dark energy (a recent phase of
accelerated expansion) could be mimicked by a deviation from standard
gravity at large scales.

To better constrain and understand the present acceleration of the
expansion, there is a crucial need for multiple observational probes.
The Integrated Sachs-Wolfe (ISW) effect (Sachs \& Wolfe 1967)
imprinted in the CMB and its correlation with the distribution of
matter at lower redshifts (through the galaxy surveys) is one of them.
The ISW effect arises from the time-variation of scalar metric
perturbations and offers a promising new way of inferring cosmological
constraints (e.g. Corasaniti, Gianantonio \& Melchiorri 2005, Pogosian
et al. 2006). It is usually divided, in the literature, into an early
ISW effect and a late ISW effect. The early effect is only important
around recombination when anisotropies can start growing and the
radiation energy density is still dynamically important. The late ISW
effect originates, on the other hand, long after the onset of matter
domination.  It is to this latter effect that we refer to here as the
ISW effect. The origin of the late ISW effect lies in the time
variation of the gravitational potential (e.g. Kofman \& Starobinsky
1985, Kamionkowski \& Spergel 1994).  In a flat universe, the
differential redshift of photons climbing in and out of the
potential is zero except in a low matter density universe and at the
onset of dark energy domination.

The ISW effect is seen mainly in the lowest $l$-values range of the
CMB temperature power spectrum ($l < 30$). Its importance comes from
the fact that it is sensitive to the amount, to the equation of state
and to the clustering properties of the dark energy. Detection of such
a weak signal is, however, limited by cosmic variance. But because the
time evolution of the potential that gives rise to the ISW effect may
also be probed by observations of large scale structure (LSS), the
most effective way to detect the ISW effect is through the
cross-correlation of the CMB with tracers of the LSS distribution.
This idea, first proposed by Crittenden \& Turok (1996), has been
widely discussed in the literature (e.g. Kamionkowski 1996,
Kinkhabwala \& Kamionkowski 1999, Cooray 2002, Afshordi 2004, Hu \&
Scranton 2004). A detection of the ISW effect was first attempted
using the COBE data and radio sources or the X-ray background (Boughn,
Crittenden \& Turok 1998, Boughn \& Crittenden 2002) without much
success. The recent WMAP data (Spergel et al. 2003, 2007) provide high
quality CMB measurements at large scales.  They were used in
combination with many LSS tracers to re-assess the ISW detection.  The
correlations were calculated using various galaxy surveys (2MASS,
SDSS, NVSS, APM, HEAO). However, despite numerous attempts in real
space (Diego, Hansen \& Silk 2003, Boughn \& Crittenden 2004, Cabre et
al. 2006, Fosalba \& Gaztanaga 2004, Hernandez-Monteagudo \&
Rubiono-Martin 2004, Nolta et al. 2004, Afshordi, Loh \& Strauss 2004,
Padmanabhan et al. 2005, Gaztanaga, Maneram \& Multamaki 2006,
Giannantonio et al. 2006, Rassat et al.  2006) or in the wavelet
domain (e.g. Vielva, Martinez-Gonzalez \& Tucci 2006, McEwen et al.
2007), the ISW effect is detected through correlations with only weak
significance. But the CMB and LSS surveys are now entering a
precision age when they can start contributing to a stronger ISW
detection, and hence provide valuable cosmological information, in
particular about dark energy.

In this work we explore the power of next generation CMB and LSS
surveys in constraining the nature of dark energy through the
cross-correlation of the ISW effect and the galaxy distribution.  We
start by using a signal to noise analysis in order to find the most appropriate
properties of a survey that will allow to detect the correlated signal at a
minimum of $4$ sigma. Then we investigate the power of a next
generation experiment, obeying the aforementioned characteristics, in
constraining different dark energy models.

In Section~\ref{sec:isw} we revise the auto- and the cross-correlation angular
power spectra of the ISW and of the galaxy distributions and model 
the different contributions entering the analysis. In
Section~\ref{sec:sn} we focus on the signal to noise analysis allowing us to
investigate the optimisation of the galaxy survey to the ISW
detection. We thus quantify the requirements for an optimal next
generation survey planned within the context of the Cosmic Vision call
for proposal, namely the DUNE mission (Refregier et al. 2006, 
{\tt http://www.dune-mission.net/}). In Section~\ref{sec:fisher}, we present a Fisher analysis to
determine the future constraints on the dark energy equation of state
through the correlation between CMB and LSS surveys.  Finally we
discuss the results and give our conclusions in Section~\ref{sec:conclusions}.

\section{Correlating the ISW effect and galaxy surveys.}
\label{sec:isw}

The ISW effect is a contribution to the CMB temperature anisotropies
that arises in the direction $\hat{n}$ due to variations of the
gravitational potential, $\Phi$, along the path of CMB photons from last
scattering until now,
\begin{equation}
 \frac{\Delta T_{\rm ISW}}{T}\left(\hat{n}\right) = - 2
\int_{0}^{r_0} dr\, \dot{\Phi}(r,\hat{n}r)\, 
\end{equation}
where $\dot{\Phi}\equiv \partial \Phi / \partial r$ can be
conveniently related to the matter density field $\delta$ through the
Poisson equation, assuming that the dark energy component does not
cluster (for a clustering model see e.g. Weller \& Lewis 2003). The
variable $r$ is the conformal distance (or equivalently conformal
time), defined today as $r_0$, and given by
\begin{equation}
r(z) = \int_{0}^{z} \frac{dz'}{H_0 E(z')}
\label{eq:r_z}
\end{equation}
where, for a $\Lambda$CDM cosmology, $E(z)^2 \equiv \Omega_{\rm
m}(1+z)^3 + \Omega_{\rm K}(1+z)^2 + \Omega_{\Lambda}$ with
$\Omega_{\rm M}$, $\Omega_{\rm K}$ and $\Omega_{\Lambda}$
corresponding to the energy density contributions of the matter, the
curvature and the cosmological constant today in units of the critical
density, and $H_0$ is the present day Hubble constant. We set $c=1$. If
the dark energy is described by a quintessence field with a present
energy density $\Omega_{\rm DE}$ and equation of state $w(z)$,
then $\Omega_{\Lambda}$ in $E(z)$ is replaced by $\Omega_{\rm
DE}(1+z)^3\exp[3\int_0^z{\rm d}z'\, w(z')/(1+z')]$.

In a flat universe ($\Omega_{\rm K}=0$), within the linear regime of
fluctuations, the gravitational potential does not change in time if
the expansion of the universe is dominated by a fluid of constant
equation of state.
Therefore, for most of the time since last scattering, matter
domination ensured a vanishing ISW contribution. Conversely, however, a
detection of an ISW effect would indicate that the effective equation
of state of the universe actually changed. This is most interesting in
particular with respect to the dark energy dominated era.

It is noteworthy that the temperature change due to the gravitational
redshifting of photons in the ISW is frequency independent and cannot
be separated from the primary anisotropies using spectral information
only.

\vspace{3mm}

Since the temperature of the CMB photons is modified, in the dark
energy dominated regime, as they traverse an over-density, the most
effective way to detect the ISW effect is through its
cross-correlation with the large scale structure distribution. We
therefore present in the following the formalism used for computing the
cross-correlation signal, as well as the auto-correlations for both 
galaxies and CMB.

\begin{figure}[!h]
   \centering
    \includegraphics[width=8cm, height=6cm]{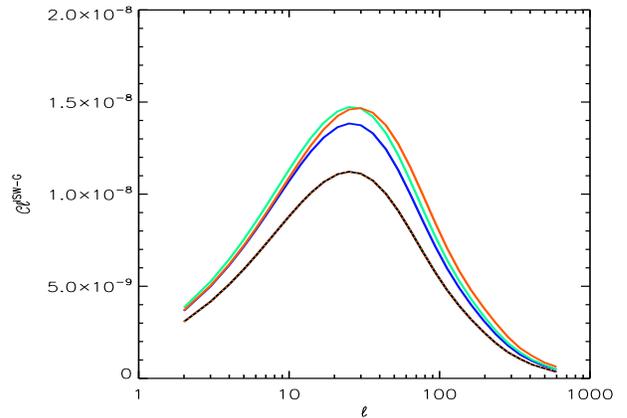}
    \caption{The angular power spectrum of the correlation between LSS
      and CMB signals is shown for different cosmologies (see
      Table~\ref{tabletheta} for details).  The $\Lambda$CDM model is
      shown in black, model A2 in dark blue, model B in green, model C
      with a transition at $a_t=0.5$ ($a_t=0.2$) in red solid (dotted)
      line.  }
\label{fig:clisw}
\end{figure}

By expanding the ISW temperature fluctuations in the sky in spherical
harmonics, it is straightforward to show that the angular power
spectrum of the ISW effect is given by (see e.g. Cooray 2002)
\begin{equation}
C_l^\isw = {2 \over \pi} \int {\rm d}k\, k^2 P_{\delta\delta}(k)
                \left[I_l^\isw(k)\right]^2 \,,
\label{eqn:clexact}
\end{equation}
where $P_{\delta\delta}(k)$ is the power spectrum of density fluctuations today, and 
\begin{equation}
I_l^\isw(k) = \int_0^{\rad_0} {\rm d}\rad\, W^\isw(k,\rad) j_l(k\rad) \, .
\end{equation}
The ISW window function, in the case of a spatially flat
Universe with non-clustering dark energy,
is 
\begin{equation}
W^\isw(k,\rad) = -3\Omega_{\rm M} \left(\frac{H_0}{k}\right)^2 \dot{F}(r) \, .
\label{eq:wisw}
\end{equation}
The $j_l$ are spherical Bessel functions of the first kind and $F(r)
\equiv G(r)/a$ is the linear over-density growth factor $G$ divided by
the scale factor $a$. $G$ relates the density field $\delta$ at any
redshift with its present day value as $\delta(k,r) = G(r) \delta(k,r=0)$
and, for a $\Lambda$CDM cosmology, is given in function of redshift by
(see Heath 1977 and Eisenstein 1997)
\begin{equation}
G(z)  \propto \Omega_{\rm M} E(z) \int_{z}^{\infty}
{\rm d}z'\,\frac{1+z'}{E^3(z')}.
\end{equation}
It is normalized such that $G(z=0)=1$. In the following, we use
Linder approximation for the growth factor (Linder
2005) which writes
\begin{equation}
G(z)  \propto \exp \left\{\int^z_\infty\left[\Omega_{\rm
      m}(z')^\gamma-1\right]\frac{{\rm d}z'}{1+z'}\right\}\,,
\label{eq:linder}
\end{equation}
where $\Omega_{\rm M}(z)=\Omega_{\rm M}\frac{(1+z)^3}{E(z)^2}$ and
$\gamma$ is a parameter set to 0.55 for the $\Lambda$CDM model. More
generally, $\gamma=0.55+0.05[1+w(z=1)]$ for an equation of state
$w>-1$ and $\gamma=0.55+0.02[1+w(z=1)]$ for $w<-1$. Equation
\ref{eq:linder} was shown to be a good approximation to the growth
factor for dark energy models with both a constant and a varying equation
of state; moreover, it approximates well the growth factor in modified
gravity models (e.g. Linder 2005, Amendola, Kunz \& Sapone 2007,
Huterer \& Linder 2007 and references therein).

\begin{figure}
 \centering
    \includegraphics[width=8cm]{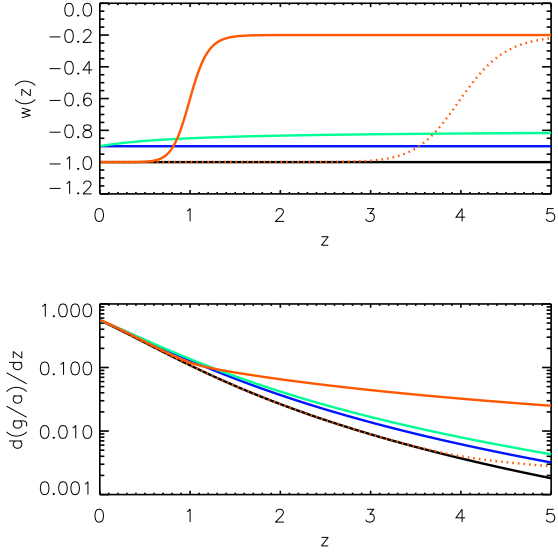}
\caption{Top: Dark energy equation of state, following parametrisation A1 (black), A2 (blue), B (green) and C (red) with the cosmological parameters given in Table \ref{tabletheta}. The dotted line corresponds to a parametrisation C with $a_t=0.2$; Bottom: corresponding $\dot{F}(z)$ (c.f. Eq.~\ref{eq:wisw})}
	\label{fig:wdez}
\end{figure} 

\begin{table}
 \begin{tabular}{|l|c|c|c|c|c|c|c|c|}
\hline
Model& $H_0$ & $\Omega_{\rm b}$& $\Omega_{\rm M}$ & $\sigma_{\rm 8}$ & $ n_{\rm s}$ & $w_0$ & $w_a$& $a_t$ \\ \hline
 A1&  73& 0.04& 0.24 & 0.74 & 0.951&  -1  & -- & -- \\ \hline
 A2&  73& 0.04& 0.24 & 0.74 & 0.951&  -0.9  & -- & -- \\ \hline
 B &  73& 0.04& 0.24 & 0.74 & 0.951&  -0.9  & 0.1 & -- \\ \hline
 C &  73& 0.04& 0.24 & 0.74 & 0.951&  --  & -- & 0.5 \\ \hline
 \end{tabular} 
\caption{Values of cosmological parameters for the fiducial models used in
the Fisher analysis. Note that we impose flatness for all models.}
\label{tabletheta}
\end{table}

In the following analysis we consider three paramerisations (A,
B, and C) of $w$, as shown in Fig.~\ref{fig:wdez}:
\begin{itemize}
\item (A) constant equation of state: $w=-1$ (A1, i.e. the $\Lambda$CDM model) and $w=-0.9$ (A2)
\item (B) linear evolution with the scale factor: $w(z)=w_0+w_a\frac{z}{1+z}$
\item (C) kink parametrisation, where the equation of state undergoes
a rapid transition at a particular redshift $z_{\rm t}$: $w(z) =
\frac{1}{2} (w_\START + w_\END) - \frac{1}{2} (w_\START - w_\END)
\tanh \left(\Gamma \log\left(\frac{1+z_{\rm t}}{1+z}\right) \right)$.
$w_\START$ and $w_\END$ are the two assymptotic values (Douspis et
al. 2006, Pogosian et al 2006, Corasaniti et al 2004).
\end{itemize}
These three parametrizations correspond to three different structure formation histories. The respective growth factor 
evolution, key ingredient of the ISW effect as shown by Eq.~\ref{eq:wisw}, is presented in the bottom panel of Fig.~\ref{fig:wdez}. 

Going back to Eq.~\ref{eqn:clexact}, we have defined the power spectrum of density fluctuations by $
\langle \delta (\vec{k}) \delta^* (\vec{k}')\rangle = (2 \pi)^3
\delta_{\rm D} (\vec{k} + \vec{k}') P_{\delta\delta}(k)$ where
\begin{equation}
P_{\delta\delta}(k) \propto 2\pi^2 \left( \frac{k}{H_0}
\right)^{n_{\rm s}+3}
\frac{\mathcal{T}^2(k)}{k^3}\,,
\end{equation}
with scalar spectral index $n_{\rm s}$. The wavenumber $k$ is
expressed throughout in $h$Mpc$^{-1}$. We use $H_0^{-1}=2997.9 \, h^{-1}$Mpc as the
inverse Hubble distance today.  For the transfer function
$\mathcal{T}$ we utilise the fitting formulae given in Eisenstein \& Hu
1997 for an arbitrary CDM+baryon universe. We use the proportionality symbol in the previous equation 
because the power spectrum is
normalized at small angular scales to the cluster abundances which
fix $\sigma_8$, the variance in the mass enclosed in spheres of
radius $R = 8 h^{-1}$Mpc. In terms of the power spectrum, we have
$\sigma_R^2=1/(2\pi)^2 \int k^2 {\rm d}k P_{\delta\delta}(k) [ 3 H_0
j_l(kR/H_0)/kR]^2$ (see eg Jaffe \& Kamionkowski 1998).

Since we are interested in the cross-correlation of the CMB and galaxy
distribution in large surveys, we define, in a similar manner, the
2-point angular cross-correlation of the ISW temperature anisotropies
with the galaxy distribution field
\begin{equation}
C_l^{\isw-{\rm G}} = {2 \over \pi} \int {\rm d}k\, k^2  P_{\delta\delta}(k)
                I_l^\isw(k) I_l^{\rm G}(k) ,
\end{equation}
where
\begin{equation}
I_l^{\rm G}(k) = \int_0^{\rad_0} {\rm d}\rad\, W^{\rm G}(k,\rad)
j_l(k\rad) \, ,
\end{equation}
and the galaxy window function is given by
\begin{equation}
W^{\rm G}(k,\rad) = b_{\rm G}(k,\rad) n_{\rm G}(\rad) G(\rad) \, .
\end{equation}
$b_{\rm G}(k,\rad)$ is the (in principle) scale- and redshift-dependent 
bias of the galaxies we consider, and $n_{\rm G}(z)=n_{\rm
G}(r)/H(z)$ is their normalised redshift distribution defined by
\begin{equation}
n_{\rm G}(z) = n_{\rm G}^0\left(\frac{z}{z_{0}}\right)^{\alpha} \exp
\left[-\left(\frac{z}{z_{0}}\right)^\beta \right]\, , \label{nGz}
\end{equation}
where the variables $\alpha$ and $\beta$ provide a description of the
galaxy distribution at low and at high redshifts respectively and
$z_{0}$ corresponds to the median redshift $z_{\rm med}\simeq 1.4
\,z_0$. The variable $n_{\rm G}^0$ is a normalization such that
$\int_{0}^{r_0} n_{\rm G}(r) dr = 1$.

Examples of the angular power spectrum of the cross-correlation between CMB and galaxy
distribution in large surveys are shown in Fig.~\ref{fig:clisw}, for the different cosmologies 
of Table \ref{tabletheta}. In our analysis we neglect both the presence of massive neutrinos (c.f. Lesgourgues, Valkenburg and Gaztanaga 2007) and magnification effects which are relevant at very high redshift (c.f. LoVerde, Hui and Gaztanaga 2007).

\section{Signal-to-Noise analysis}
\label{sec:sn}

As a first step, we start by investigating the detection level of the
ISW effect in cross-correlation. To do this, we perform a signal-to-noise (SN)
analysis. Using the power spectra computed in the previous section we
can write the total signal-to-noise of the ISW detection as (Cooray 2002,
Afshordi 2004):
\begin{eqnarray}
&& \left(\frac{\rm S}{\rm N}\right)^2 = f^c_{\rm sky} \sum_{l=l_{\rm
min}}^{l_{\rm max}} (2l+1) \nonumber \\ &\times&
\frac{\left[C_l^{\isw-{\rm G}}\right]^2}{\left[C_l^{\isw-{\rm
G}}\right]^2 + \left(C_l^\isw+N_l^{\rm ISW}\right)\left(C_l^{\rm
G}+N_l^{\rm G}\right)} \, ,
\label{eqn:sn}
\end{eqnarray}
where $f^c_{\rm sky}$ is the fraction of sky common to the CMB and the
galaxy survey maps, and the total (or cumulative) signal-to-noise is
summed over multipoles between $l_{\rm min}=2$ and $l_{\rm max}=60$
(where the signal has its major contribution).  The spectra
$C_l^{\isw-{\rm G}}$ and $C_l^\isw$ were defined in the previous
section, and $C_l^{\rm G}$ is straightforward to obtain.  The noise
contribution in the ISW signal and galaxy surveys are $N_l^{\rm ISW}$
and $N_l^{\rm G}$ respectively. At the scales of interest for the ISW
detection, and for the CMB experiment considered, the ISW noise is
defined as $N_l^{\rm ISW} = C_l^{\rm CMB} + N_l^{\rm CMB exp}$ where
$N_l^{\rm CMB exp}$ is the contribution from the experimental
noise. The previous expression is dominated by the sample
variance. However, we do include $N_l^{\rm CMB exp}$ in our
calculations (as modeled in Knox 1995) for an experiment such as the
Planck satellite (see Table \ref{tb:lss} for the CMB noise
characteristics).  The galaxy survey noise is defined by the shot
noise contribution: $N_l^{\rm G} = \frac{1}{\bar{N}}$ where $\bar{N}$
is the surface density of sources per steradian that one can
effectively use for the correlation with CMB temperature data.  The
noise part of the SN depends then, at first order, on the common sky
fraction, on the surface density of sources, and on their median
redshift through the amplitude of the cross-power spectrum
$C_l^{\isw-{\rm G}}$ and of the galaxy auto-correlation signal
$C_l^{\rm G }$. With this analysis we are then able to optimise the
forthcoming galaxy surveys for an ISW detection in cross-correlation
as a function of their three main parameters namely $z_{\rm med}$,
$f^c_{\rm sky}$ and $\bar{N}$. We use the cosmological parameters
values from WMAP3 (Spergel et al 2007) as listed in table
\ref{tabletheta}.

\begin{figure}[!t]
   \centering
    \includegraphics[width=4cm]{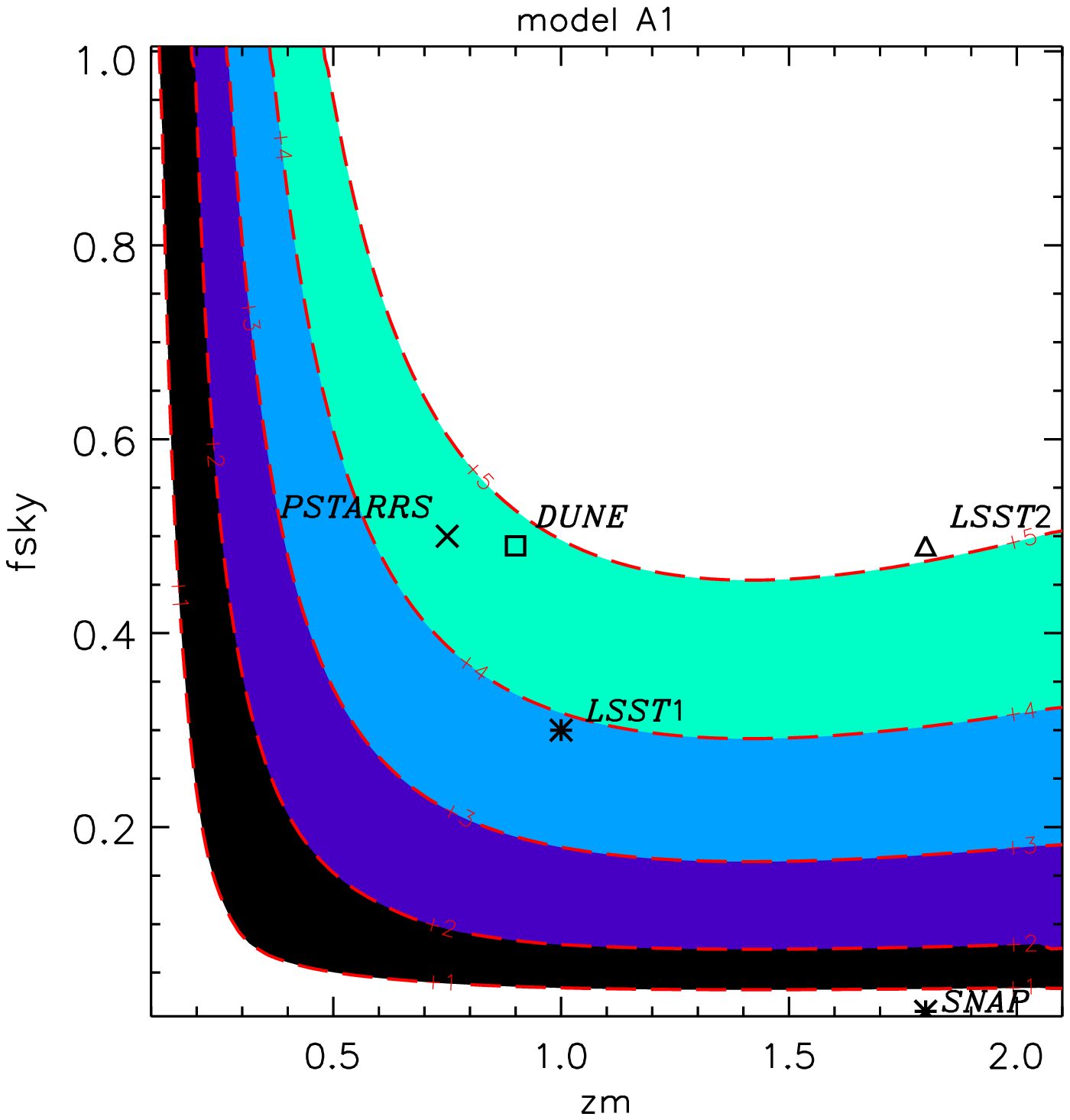}
    \includegraphics[width=4cm]{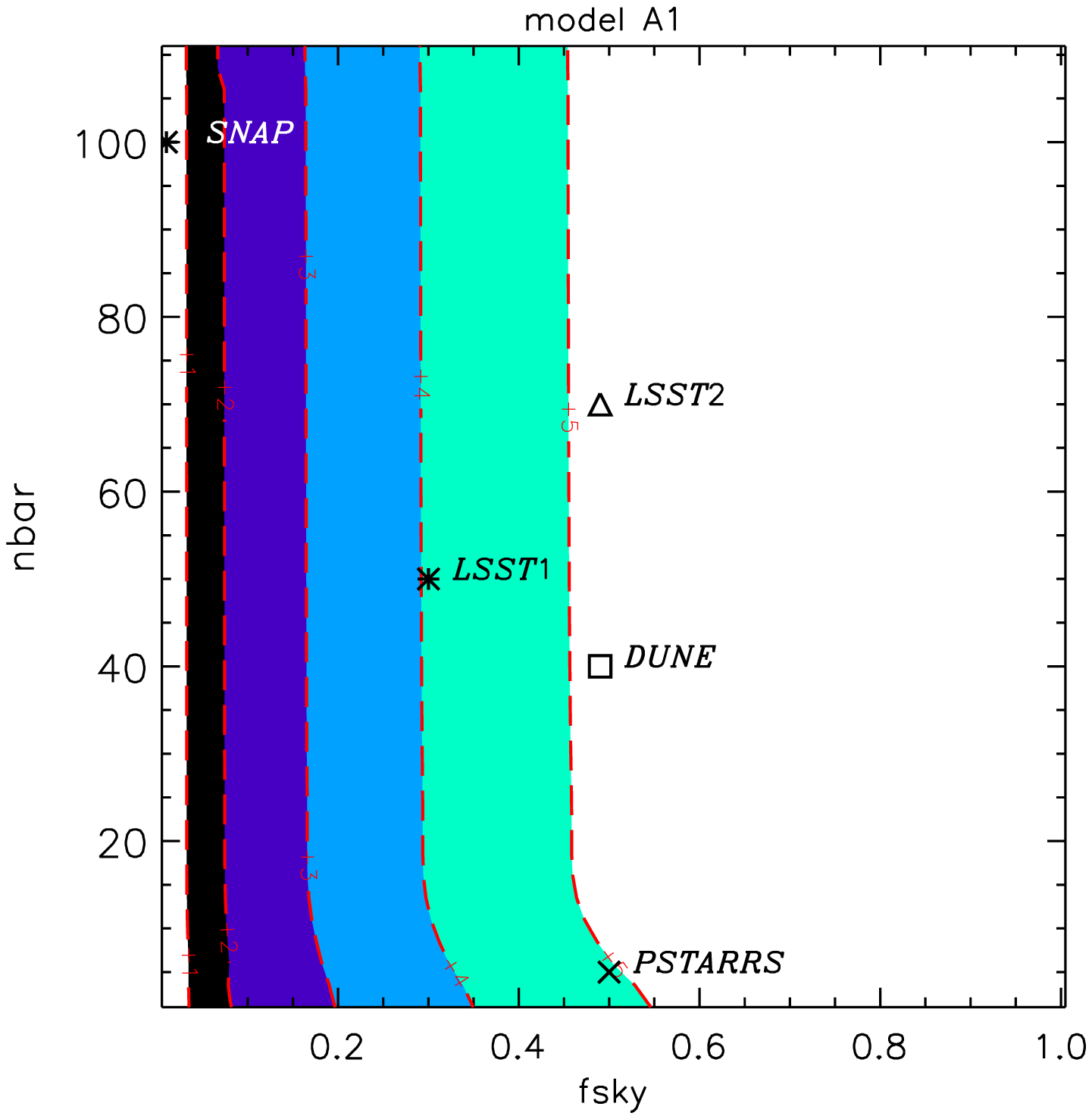}
    \caption{\label{SNfig-lambda} Total signal-to-noise for a ISW detection in
    the $\Lambda$CDM model (parametrisation A1) as a function of the galaxy
    survey parameters $f^c_{sky}$, $\bar N$ in units of arcmin$^{-2}$
    and $z_{\rm med}$. The different colours show different levels of 
    detection in number of sigmas.}
\end{figure}

\begin{figure}[!t]
   \centering
    \includegraphics[width=4cm]{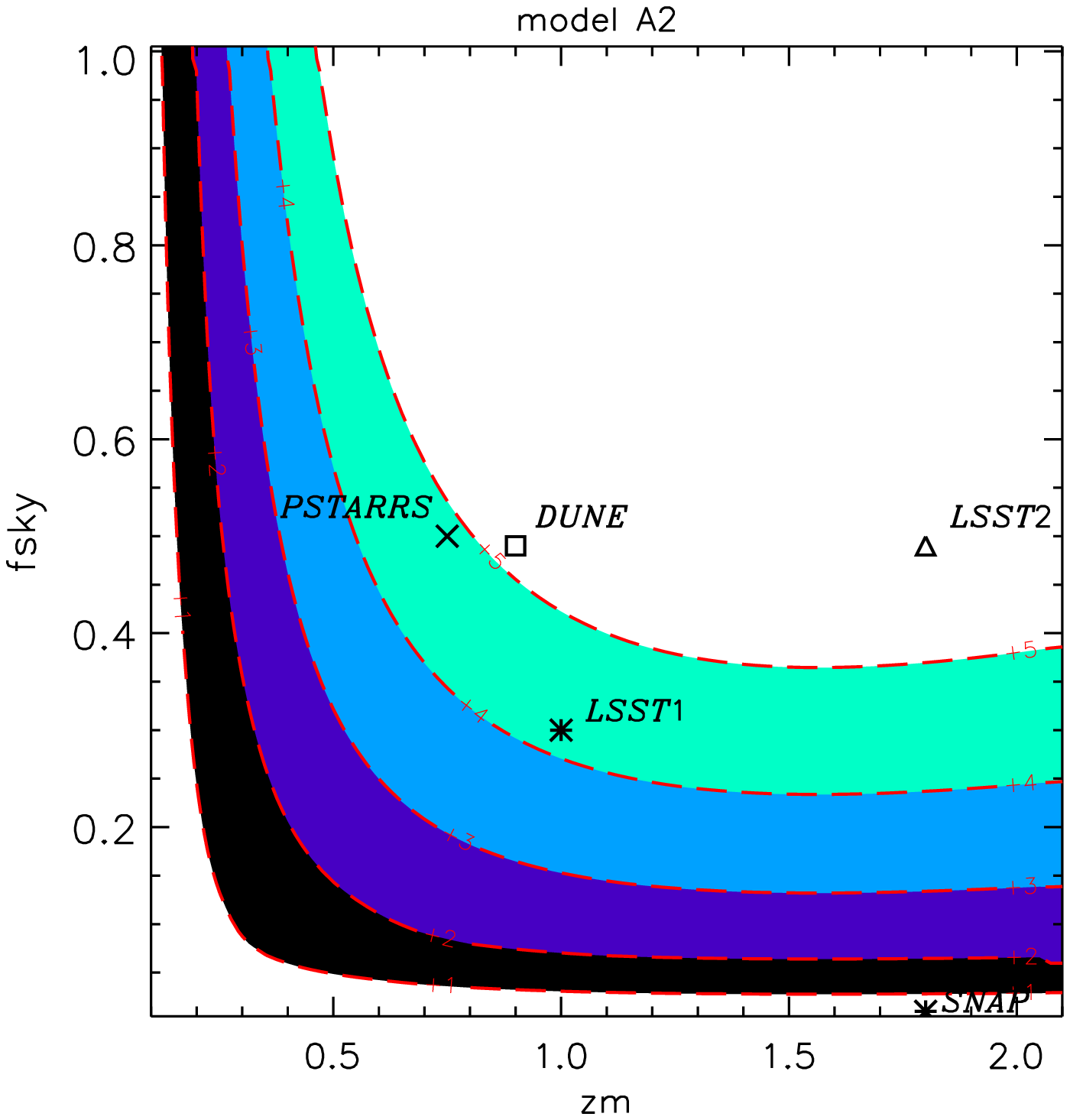}
    \includegraphics[width=4cm]{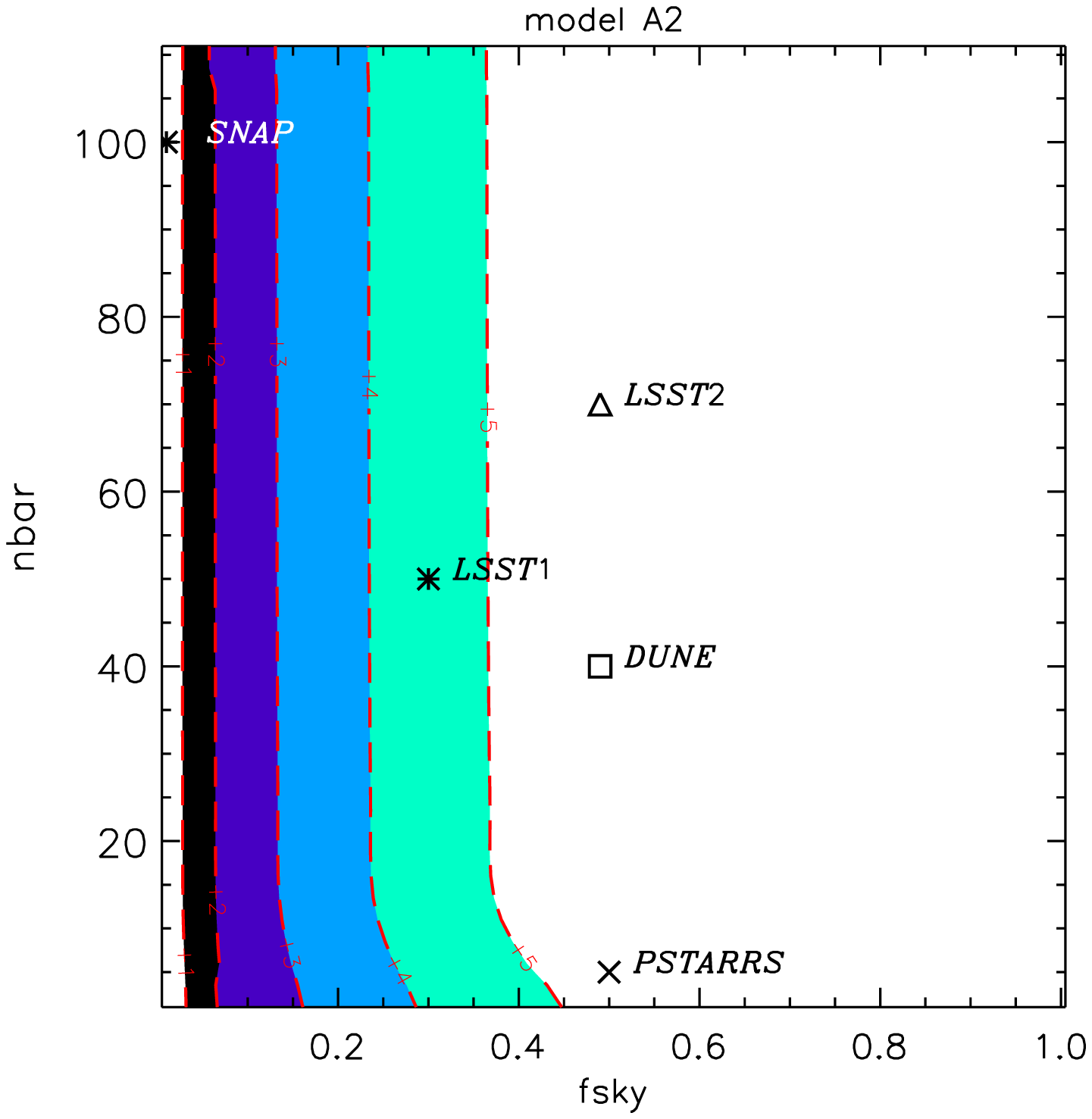}
    \caption{\label{SNfig-w09} Total signal-to-noise for a ISW detection in
    the constant equation of state model with $w=-0.9$ (parametrisation A2) as
    function of the galaxy survey parameters $f^c_{sky}$, $\bar N$ in
    units of arcmin$^{-2}$ and $z_{\rm med}$.}
\end{figure}

\begin{figure}[!t]
   \centering
    \includegraphics[width=4cm]{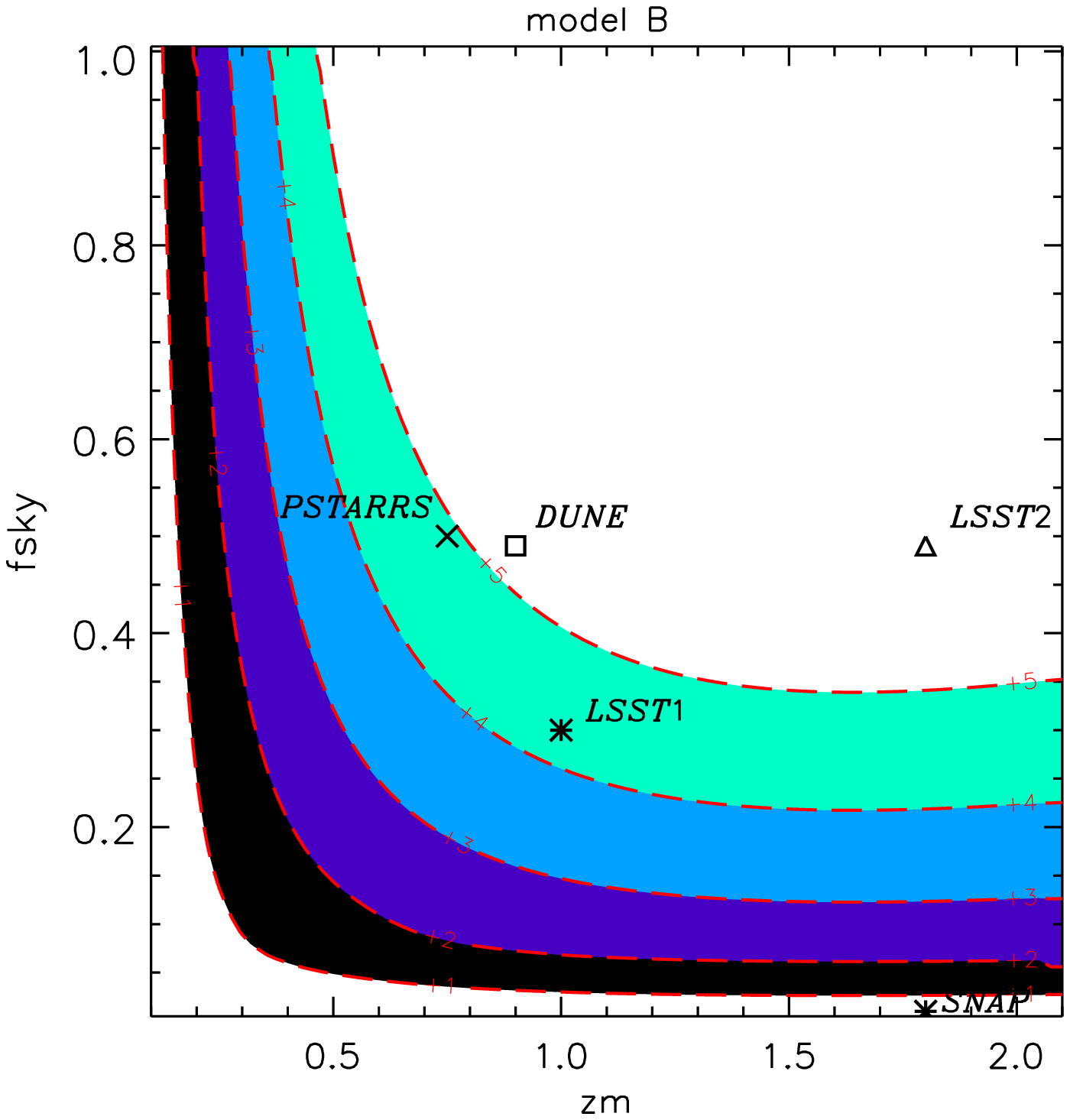}
    \includegraphics[width=4cm]{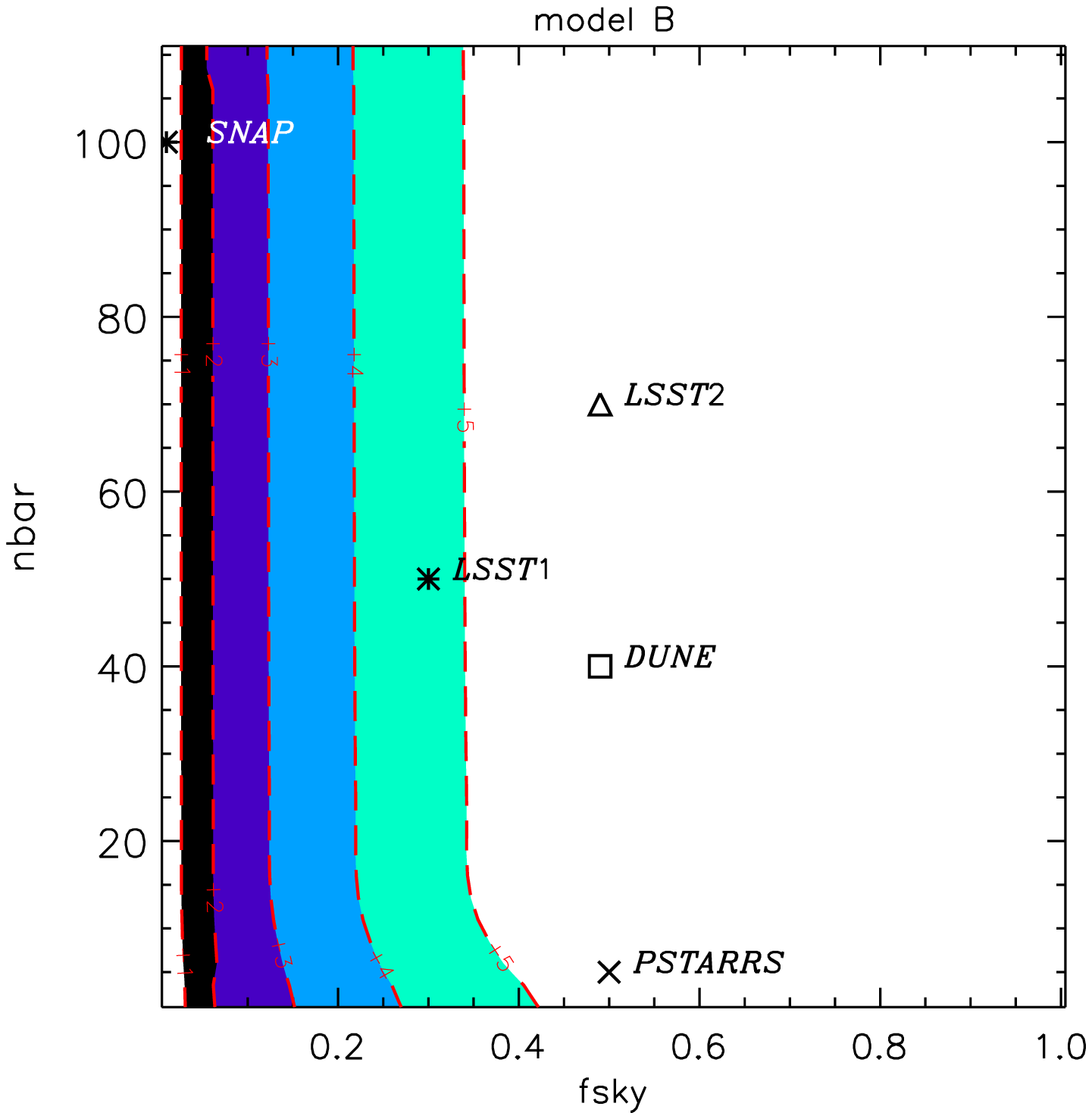}
    \caption{\label{SNfig-wz} Total signal-to-noise for a ISW detection in the
   linearly varying equation of state model (parametrisation B) as
    function of the galaxy survey parameters $f^c_{\rm sky}$, $\bar N$
    in units of arcmin$^{-2}$ and $z_{\rm med}$.}
\end{figure}

We explore the 3D parameter space, and in Figs.~\ref{SNfig-lambda},
~\ref{SNfig-w09} and~\ref{SNfig-wz} we show the SN values in 2D
diagrams where the third parameter, respectively $\bar N$ for the left
panel and $z_{\rm med}$ for the right panel, has been marginalised
over. In order to have an insight on the detection level of the ISW,
we have computed the SN values for the dark energy models A and B
given in Sect. \ref{sec:isw}. We do not consider the kink
parametrisation (model C) as it gives very similar results to the
linear parametrisation. All the results shown in this section were
obtained with a redsfhit and scale independent bias $b_{\rm G}=1$ and
with the parameters $\alpha=2$ and $\beta=1.5$ for the galaxy redshift
distribution. This set of parameters is typical for optical galaxies
studies (Heavens et al. 2007).

From all these figures, we can see that once the number density of
observed sources $\bar{N}$ reaches a given value $\bar{N}_{\rm lim}$
(typically about 10 sources per arcmin$^2$ or a bit less for all dark
energy models), the SN is constant and independent of $\bar{N}$. This
can be understood by going back to the definition of the survey noise
($N_l^{\rm G}$) entering the SN ratio equation
(Eq.~\ref{eqn:sn}): in this regime, the contribution from the Poisson noise becomes negligible. 
As a result, for an equal sky fraction $f^c_{\rm sky}$ and
median redshift $z_{\rm med}$, all surveys satisfying the condition
$\bar N > \bar{N}_{\rm lim}$ will give equivalent ISW detections.  At
a fixed $\bar N> \bar{N}_{\rm lim}$, the SN ratio is on the contrary
very sensitive to $f^c_{\rm sky}$ and $z_{\rm med}$. The former
entering the SN through the CMB noise as a multiplicative factor, it is
obvious that, for a given $z_{\rm med}$, the larger $f^c_{\rm sky}$ the
higher the detection level. Conversely, at a given $f^c_{\rm sky}$,
increasing $z_{\rm med}$ significantly improves the ISW detection only up
to $z_{\rm med}\sim 1$. In the chosen dark energy model, in fact, dark energy domination always
occurs at $z<1$. 

More specifically, in the $\Lambda$CDM model a detection of the ISW
signal with a confidence of 4$\sigma$ is attained for median redshifts
$z_{\rm med}> 0.84$ and fractions of sky $f^c_{\rm sky} > 0.35$. For a
constant equation of state model ($w=-0.9$), we find a slightly lower
median redshift, of $\sim 0.8$, and a slightly lower sky fraction
$f^c_{\rm sky}\sim 0.33$, and the same numbers apply for the varying
equation of state model. The small increase in SN in models A2 and B
is expected, since the ISW is an integrated effect. In the last two
models, dark energy domination occurs earlier, and structures grow
faster. Therefore, they give a stronger contribution to the ISW effect
than $\Lambda$CDM, providing a better SN for lower median redshift.

In the context of the dark energy models used here, we conclude that
in order for a galaxy survey to enable a detection of the ISW effect
in cross-correlation with a high enough signal-to-noise ratio it is
sufficient to design it based on the predictions from the $\Lambda$CDM
model.  The $\Lambda$CDM model gives, in fact, the most conservative
detection levels. An optimal survey (with a detection level above 4
$\sigma$) should thus be designed so that it has a minimum number
density of sources of about around 10 galaxy per arcmin$^{-2}$, covers
a sky fraction of at least 0.35 and is reasonably deep, with a minimum
median redshift larger than 0.8. One of the surveys satisfying such
conditions and being planned is the DUNE mission proposed to the ESA's
Cosmic Vision call for proposal (Refregier et al. 2006, 2008 {\tt
  http://www.dune-mission.net/}). It will provide a detection of
almost 5 sigmas, as shown in the 2D figures together with other future
galaxy surveys such as SNAP, PanSTARRS and LSST (see
Figs.~\ref{SNfig-lambda},~\ref{SNfig-w09}~and~\ref{SNfig-wz} and Table
\ref{tb:lss}).

\begin{table}[h]
\begin{center}
\begin{tabular}{|c|c|c|c|}
\hline
 & $f_{\sky}$&$z_{\rm med}$&nbar (arcmin$^{-2})$\\
\hline
\hline
DUNE(1)& 49\%  & 0.9 & 40 \\
\hline
LSST-1(2) & 30\%  & 1 & 50     \\
\hline
LSST-2(3) & 49\%  & 1.8 & 70     \\
\hline
SNAP(4) & 1 \% & 1.8 & 100    \\
\hline
PanSTARRS(5) & 50\%  &  0.75  & 5   \\
\hline
\hline
PLANCK & $f_{\rm sky}=80\%$ & $\theta_{\rm beam}=7{\rm arcmin}$& $\omega^{-1}_T = 4e^{-17} $ \\
\hline
\end{tabular}
\end{center}
\caption{\label{tb:lss} Future LSS surveys characteristics (1) from
  Refregier et al. 2006, 2008 and the DUNE website, (2) and (3) from
  Pogosian et al. 2005 as ``conservative'' and ``goal'' cases
  respectively, (4) from the SNAP collaboration, the SNAP website {\tt
    http://snap.lbl.gov/} and Aldering et al 2007, (5) from Stubbs et
  al 2007, Heavens et al. 2007 and private communications with
  S. Phleps.  Planck characteristics used for the noise part of the
  signal--to--noise, and for the Fisher matrices analyses are also
  given. The values of $z_{\rm med}$ for SNAP and $\bar N$ for
  PanSTARRS are only indicative.}
\end{table}

\section{Fisher Matrix analysis}
 \label{sec:fisher}

In order to quantify the constraint that the cross-correlation of a
next generation large scale survey with CMB maps would give on dark
energy through the ISW signal we perform a Fisher matrix analysis.
For the ISW measurement, such an analytical approach has been shown to
yield very accurate error bars by comparison with realistic
Monte-Carlo simulations of CMB and galaxy maps by Cabr\'e et al. (2007).
We use this technique to compute the errors on a set of cosmological
parameters $\Theta$. We assume the usual experimental characteristics,
such as the noise and the sky fraction, of Planck and DUNE surveys as
listed in Table \ref{tb:lss}, as these are excellent examples of the
next generation of CMB and LSS experiments. They ensure a good SN
detection as demonstrated in the previous section.

Given the characteristics of the CMB and the LSS experiments, a
fiducial model, and a cosmological framework, the smallest possible
errors on a set of parameters when determined jointly were shown to be
given by the Fisher matrix $F$ elements: $\delta \Theta_i =
\sqrt{(F^{-1})_{ii}}$ (see e.g. Tegmark, Taylor and Heavens 1997).  In
our case, the cross-correlation Fisher matrix for parameters
$\Theta_i$ and $\Theta_j$ is given by
\begin{equation}F^{i,j} = f^c_{sky}\sum_l (2l+1)
\frac{\partial C^{\isw-\X}_l}{\partial \Theta_i} \; cov^{-1}(l) \;
\frac{\partial C^{\isw-\X}_l}{\partial \Theta_j}
\end{equation}
where \begin{equation}
\label{eq:covfish}
cov(l) = {\left[C_l^{\isw-\X}\right]^2 +
\left(C_l^\isw+N_l^{\rm CMB}\right)\left(C_l^{\X}+N_l^{\rm
G}\right) }\,.
\end{equation}
The summation is done over the range of multipoles $\sim
\pi/(2f^c_{sky}) < l < 800$. The covariance term, as well as the
partial derivatives are evaluated at the fiducial model. See
Section~\ref{sec:sn} (Eq.~\ref{eqn:sn}) for more details about the
various terms entering  the expression.

We assume a flat universe with adiabatic scalar perturbations, a nearly
scale invariant initial power spectrum, containing baryonic and cold
dark matter, and dark energy. We assume zero curvature since if it
were not the case, dark energy would not be distinguishable from the
curvature through the ISW effect (Kunz 2007, Clarkson, Cort\^es \&
Bassett 2007). The Fisher analysis is then done on the following
cosmological parameters: $ \Theta=(H_0, \Omega_{\rm b}, \sigma_{\rm 8}, n_{\rm s}, \Omega_{\rm DE})$. In addition,
with respect to the dark energy component we study the three scenarios
A, B, and C given in Sect.~\ref{sec:isw} and summarised in Table
\ref{tabletheta}.

We also compute the Fisher matrix corresponding to the constraints
imposed by the CMB alone for the same three scenarios. We take into
account only one channel, in temperature, following the Planck
characteristics listed in Table \ref{tb:lss}. When combining the
constraints from the CMB alone and from the cross-correlation, we
consider that the experiments are independent and thus add the
corresponding Fisher matrices.

\begin{figure}
 \centering
    \includegraphics[width=8cm]{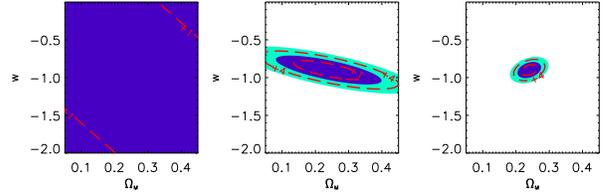}
\caption{Two-dimensional marginalised confidence intervals on the plane $(\Omega_M,w)$ obtained
with a Fisher matrix analysis centered on model A2 with $w=-0.9$. Left:
ISW constraints; center: CMB (temperature) constraints; right:
combined constraints (see text for details).}
	\label{fisher_M2}
\end{figure}

Fig.~\ref{fisher_M2} shows the constraints obtained from the Fisher
analyses of model A2 with $w=-0.9$ from the cross-correlation between
the CMB and the LSS (left), from the CMB temperature anisotropies
alone (middle) and from the combined analysis of both (right). The
panels show the confidence intervals that one could obtain on
$\Omega_{M}$ and $w$ when other parameters have been marginalised
over. As shown in the figure, the constraints from the
cross-correlation itself are quite weak, but they play a non obvious
and not negligible role in the combination. This is mainly due to the
6--dimensional shape of the likelihood and its degeneracies.

\begin{figure}
 \centering
    \includegraphics[width=8cm]{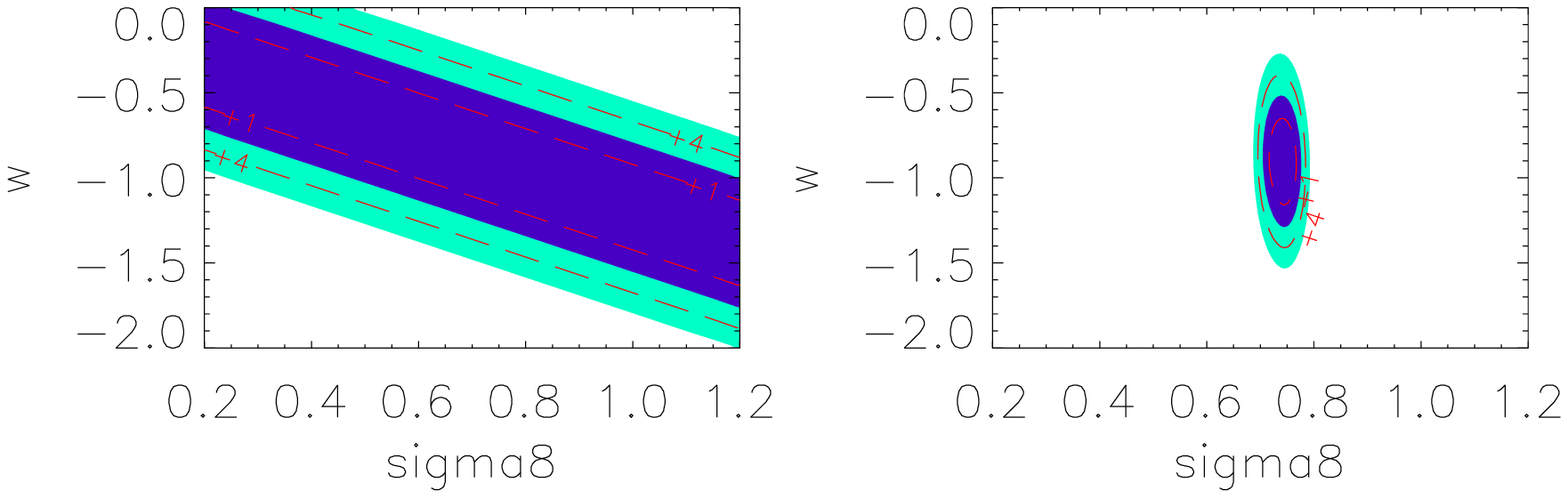}
    \includegraphics[width=8cm]{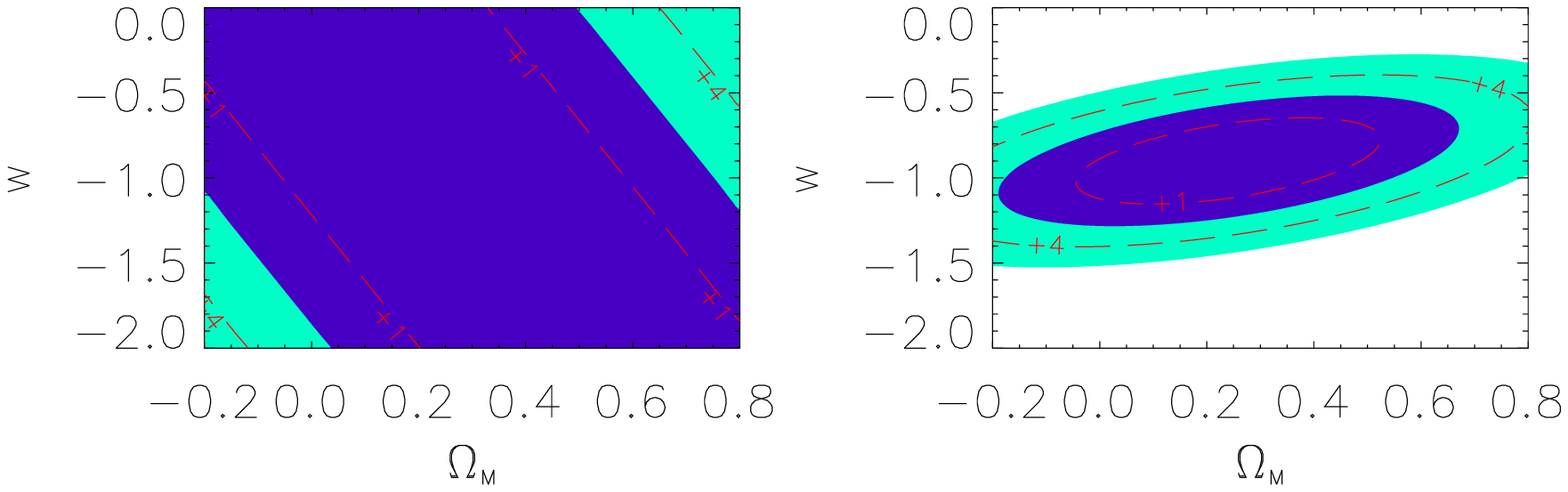}
\caption{Two-dimensional marginalised confidence intervals obtained with a Fisher matrix analysis
centered on model A2 with $w=-0.9$ for different combinations of
parameters: $(\Omega_M,w)$ and $(\sigma_8,w)$. Left: ISW constraints;
right: ISW constraints obtained with a strong prior on $\sigma_8$ (see text
for details). Note the different scales for $\Omega_M$ as compared to Fig.~\ref{fisher_M2}.}
	\label{fisher_M2+P}
\end{figure}

In order to further investigate such a result, we added strong priors
on some cosmological parameters to the cross-correlation Fisher matrix
(instead of the CMB Fisher matrix). We found that adding a prior on
the Hubble constant ($H_0$), the matter content ($\Omega_M$), or the
spectral index ($n_s$) does not improve the constraints by
much. However, the errors on the dark energy parameters are
significantly reduced when adding a prior on the amplitude of the
fluctuations via $\sigma_8$. Fig.~\ref{fisher_M2+P} shows the
constraints from the cross-correlation obtained in this last case for
scenario A2 ($w=-0.9$) without (left plots) and with (right plots) a
strong prior on $\sigma_8$ (such that $\sigma_8=0.7 \pm 0.02$).  The
top left panel shows the degeneracy between $\sigma_8$ and $w$ and
explains why, by constraining strongly $\sigma_8$, with a prior (right
panel) or with the CMB (Fig.~\ref{fisher_M2}), the equation of state
is better determined. This and other minor degeneracy breakings in the
6-dimensional space allows the ISW effect, through cross-correlation,
to improve the constraints on dark energy.

\begin{figure}
 \centering
    \includegraphics[width=8cm]{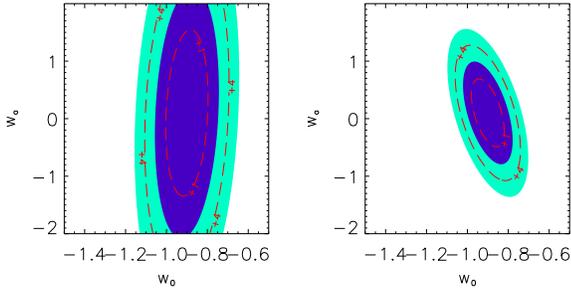}
\caption{Two-dimensional marginalized  confidence intervals obtained with a Fisher matrix analysis
for $(w_0,w_a)$ centered on model B with $w(a)=-0.9+0.1(1-a)$. Left: CMB (temperature)
constraints; right: CMB+ISW combined constraints (see text for
details).}
	\label{fisherM3}
\end{figure}

As shown previously, the CMB by itself is only able to constrain one
parameter of the dark energy model, since it is sensitive mainly to
the distance to the last scattering surface (see e.g. Pogosian et al
2006, Douspis et al 2008 and references therein). Therefore, in
scenario B with $w(a)=w_0 + (1-a) w_a$, only the value of the equation
of state at present $w_0$ is constrained (Fig.~\ref{fisherM3}
left). Once again, adding the information coming from the
cross-correlation between Planck and DUNE improves slightly the errors
on the linear expansion factor $w_a$ (right panel of
Fig.~\ref{fisherM3}). However, it does not help to distinguish a
constant and a linear dark energy model.

\begin{figure}
 \centering
    \includegraphics[width=8cm]{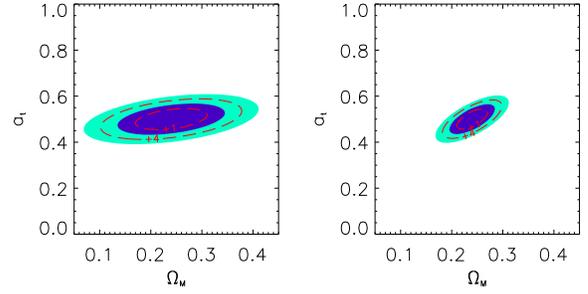}
\caption{Two-dimensional marginalized  confidence intervals obtained with a Fisher matrix analysis
centered for $(\Omega_M,a_t)$ on model C with $w(0)=-1$, $w(\infty)=-0.2$ and $a_t=0.5$; Left:
CMB (temperature) constraints; right: combined constraints (see text
for details).}
	\label{fisher_M5}
\end{figure}

Finally we study the ``kink'' model C in order to assess the
sensitivity of the cross-correlation to probe a sharp transition in
the evolution of $w$. The model considered shows a sharp transition
($\Gamma=10$) between $w(z=0)=-1$ and some arbitrary value close to 0
far in the past, e.g. $w(z=\infty)=-0.2$. Since with the CMB one can
constrain only one dark energy parameter, in this case we choose to
let free the transition redshift $z_t$. For transitions taking place
early enough in time, the equation of state is  $w=-1$ for
most of the period of structure formation. Little difference is then
expected between such a model and a $\Lambda$CDM model (see
Fig.~\ref{fig:wdez}). For recent transitions, on the contrary, huge
effects are expected.  This model has been already investigated in
Douspis et al (2008) with current CMB and SNIa data, showing that a
transition at $z_t>0.5$ ($a_t <0.66$) is allowed.  The Fisher matrix
analysis, which gives the smallest possible error bar on the
parameters, relies by construction on the hypothesis of a Gaussian
likelihood for the $C_{\ell}$. This prevents us to have asymmetric
error bars on the parameters, and in this particular case does not
allow us to obtain the realistic constraints that one could have:
i.e., that transitions at $a_t<0.4$ are also allowed. This can be seen
by comparing Fig.~\ref{fisher_M5} (left panel) with Fig.~5 of Douspis
et al 2008. Moreover, we choose to take as reference model for
scenario C (see Table \ref{tabletheta}) a reasonable value for the
period of transition, $a_t=0.5$. Due to the big difference of impact
on the growth of structure as a function of $a_t$ (see
Fig.~\ref{fig:wdez}), the errors on the transition period are also
strongly dependent on the reference model chosen.  In our case, we see
in the right panel of Fig.~\ref{fisher_M5} that adding the ISW
information to the CMB temperature anisotropy improves the constraints
on $\Omega_M$ in the plane $(\Omega_M,a_t)$, but does not improve
constraints on $a_t$.

\section{Conclusions}
\label{sec:conclusions}

In this work we have analysed the cross-correlation between the ISW
effect and a galaxy survey characterised by the redshift distribution
given by the Eq.~\ref{nGz} and assuming simple Poisson noise. We have
relied on Linder approximation given in Eq.~\ref{eq:linder} to model
the growth of structure in dynamical dark energy models.

We have determined the most appropriate properties of a survey in
order to detect an ISW signal in cross-correlation CMB/LSS at a
minimum of $4$ sigma. To do this we have used a signal to noise
analysis. Our results agree with those obtained by Afshordi (2004): we
have shown that the necessary properties for a survey to be
significantly successful in the quest for an ISW signal are a minimum
number density of sources of about 10 galaxies per arcmin$^{2}$, a
minimum sky fraction of the order of 0.35 and a minimum median
redshift of about 0.8. We indicate the DUNE project as a promising
candidate for providing a good ISW detection, once correlated with
Planck data. Furthermore, the number of galaxies detected by such a
survey is high enough to divide the distribution in different redshift
bins, allowing to probe the dark energy at different epocs. As found
above, the Poisson noise is negligible for a number of detected
galaxies which is higher than 10 per arcmin$^{-2}$. If this condition
is met in each redshift bin, this increases the signal to noise
correspondingly to the number of bins considered.

We then investigated the power in constraining different dark energy
models of typical next generation experiments having the aformentioned
characteristics.  We took the DUNE and the Planck surveys.  Here
again, we confirm the result of previous analyses, such as those of
Pogosian et al (2006) and Douspis et al (2008).  We showed that the
ISW effect does help (as compared to CMB alone) in breaking
degeneracies among the parameters describing the dark energy model and
the other cosmological parameters, primarily $\sigma_8$. The
cross-correlation allows us to put a constraint of the order of $10$\%
on $w$ for a model with a constant equation of state (A) and reduces
the errors on the estimation of the parameter $w_a$ in a linear model
(B). However, it does not permit to distinguish between a constant and
a dynamical equation of state for the dark energy. Fitting a dark
energy model with a kink, we have found that adding the
cross-correlation does not improve the CMB constraints on the
transition redshift: the ISW is therefore insensitive to sharp
transitions, and a transition at any redshift larger than $0.5$ is
still allowed.

\begin{acknowledgements}

NA and MD thank the collaboration programme PAI-PESSOA for partial
funding. They further thank Instituto Superior T\'ecnico (IST) for hospitality. PGC
is funded by the {\it Funda\c{c}\~ao para a Ci\^encia e a Tecnologia} and
wishes to thank the Institut d'Astrophysique Spatiale (IAS) for its welcoming and support, 
and Stefanie Phleps for useful conversations. CC acknowledge support by the ANR 
funding PHYS@COL\&COS, and thanks IAS and IST for hospitality. We thank Mathieu Langer for helpful comments. 

\end{acknowledgements}

\end{document}